\RequirePackage[l2tabu, orthodox]{nag}
\RequirePackage{snapshot}

\documentclass[10pt,twocolumn]{extarticle}
\makeatletter
\if@twocolumn
  \usepackage[dvips,letterpaper,top=0.5in, bottom=0.5in, left=0.75in, right=0.5in,includefoot,heightrounded]{geometry}
\else
  \usepackage[dvips,letterpaper,margin=1in,includefoot,heightrounded]{geometry}
\fi

\usepackage{srcltx}

\usepackage{amsmath}
\usepackage{amssymb,amsfonts}

\usepackage{abstract}

\usepackage{epsfig}
\usepackage[usenames,dvipsnames,svgnames,x11names]{xcolor}
\usepackage{subfigure}

\usepackage{booktabs}

\usepackage{setspace}
\usepackage{flushend}

\usepackage{cite}
\usepackage{bm}

\usepackage{enumerate}

\usepackage[short,12hr]{datetime}

\usepackage{hyperref}
       \usepackage{fouriernc}

\newcommand{\cas}{\operatorname{cas}}
\newcommand{\kernel}{\operatorname{ker}}

\newcommand{\mse}{\operatorname{MSE}}

\makeatletter

\newcommand{\printtitle}{%
\makeatletter
\if@twocolumn

\twocolumn[%
  \maketitle
  \begin{onecolabstract}
    \myabstract
  \end{onecolabstract}
  \begin{center}
    \small
    \textbf{Keywords}
    \\\medskip
    \mykeywords
  \end{center}
  \bigskip
]
\saythanks
\else
  \maketitle
  \begin{onecolabstract}
    \myabstract
  \end{onecolabstract}
  \begin{center}
    \small
    \textbf{Keywords}
    \\\medskip
    \mykeywords
  \end{center}
  \bigskip
  \onehalfspacing
\fi
\makeatother
}

\title{%
An Integer Approximation Method for Discrete Sinusoidal Transforms}

\author{%
R. J.~Cintra%
\thanks{%
R. J. Cintra is with the
Signal Processing Group,
Departamento de Estat\'istica,
Universidade Federal de Pernambuco.
E-mail: \url{rjdsc@de.ufpe.br}}
}

\date{}

\newcommand{\myabstract}{%
Approximate methods have been considered as a means to
the evaluation of discrete transforms.
In this work,
we propose and analyze a class of integer transforms
for the discrete Fourier, Hartley, and cosine transforms (DFT, DHT, and DCT),
based on simple dyadic rational approximation methods.
The introduced method is general,
applicable to several block-lengths,
whereas
existing approaches are usually dedicated to specific transform sizes.
The suggested approximate transforms enjoy
low multiplicative complexity and the orthogonality property is achievable
via matrix polar decomposition.
We show that the obtained transforms are competitive with
archived methods in literature.
New 8-point square wave approximate transforms
for the DFT, DHT, and DCT are also introduced as
particular cases of the introduced methodology.
}

\newcommand{\mykeywords}{%
Approximate transforms
discrete sinusoidal transforms,
low-complexity transforms,
nonorthogonal transforms,
orthogonalization
}

\begin{document}

\printtitle

\section{Introduction}

Discrete transforms play a significant role in digital signal processing.
Among the possible discrete transforms,
those based on sinusoidal transformation kernels
occupy a prominent position.
Examples of discrete sinusoidal transforms include
the discrete Fourier transform (DFT),
the discrete Hartley transform (DHT),
and
the discrete cosine transform (DCT)~\cite{briggs1995manual,britanak2007discrete}.

Mathematically,
discrete sinusoidal transforms
relate
two $n$-dimensional vectors
$\mathbf{v}$ and $\mathbf{v}'$
possibly
defined
over the complex numbers field
according to the
formalism below:
\begin{equation}
\label{forward_kernel}
v_k' = \sum_{i=0}^{n-1} v_i \cdot \kernel\left( i,k,n \right),
\quad
k=0,1,\ldots, n-1,
\end{equation}
\begin{equation}
\label{inverse_kernel}
v_i =
\sum_{k=0}^{n-1} v_k' \cdot \kernel^{-1}\left(  i,k,n \right),
\quad
i=0,1,\ldots, n-1,
\end{equation}
where $\ker(\cdot,\cdot,\cdot)$ and $\ker^{-1}(\cdot,\cdot,\cdot)$
are known as the forward and inverse transformation kernels, respectively.
Table~\ref{table.kernels}
lists some possible kernel functions;
the Kronecker delta was employed.

Although a variety of factors
contribute to
the computational complexity of a numerical method~\cite{briggs1995manual},
the number of required arithmetical operations
is frequently
utilized as a measure of complexity.
When computed directly
according to Equations~(\ref{forward_kernel}) and~(\ref{inverse_kernel}),
discrete sinusoidal transforms require
a number of multiplications and additions
in~$O(n^2)$.
Thus,
direct computation may not be practical.

Additionally,
kernels can be complex-valued,
which require further arithmetical considerations.
The DFT kernel is a notable example.
Moreover,
the necessary values are often irrational numbers.
Therefore,
computations described in
Equations~(\ref{forward_kernel}) and~(\ref{inverse_kernel})
not only
require a significant amount of operations,
but they are expected to handle
floating-point representation
over a possibly complex field.

\begin{table}
\centering
\caption{Some sinusoidal unitary kernels}
\label{table.kernels}
\begin{tabular}{cc}
\noalign{\smallskip}
\hline
\noalign{\smallskip}
Transform kernel
          &  $\ker(i,k,n)$ \\
\noalign{\smallskip}
\hline
\noalign{\smallskip}
Fourier   &  $\frac{1}{\sqrt{n}}\exp(-2\pi j ik/n)$
\\
\noalign{\smallskip}
Cosine    &  $\left(1 - (1-1/\sqrt{2})\delta_{0,k} \right)\sqrt{2/n}\cos(\frac{\pi (i+1/2)k}{n})$
\\
\noalign{\smallskip}
Hartley   &  $\frac{1}{\sqrt{n}}\cas(2\pi ik/n)$
\\
\noalign{\smallskip}
\hline
\end{tabular}
\end{table}

Fast algorithms constitute a collection of methods
aiming at dramatically reducing arithmetical complexity figures.
For discrete transforms,
classical methods usually
attain such minimization
by means of
(i) divide-and-conquer strategies~\cite{blahut1985fast};
(ii) matrix factorization schemes~\cite{vanloan1992frameworks};
and
(iii) convolution methods~\cite{burrus1985algorithms}.

Efforts has been directed to the reduction of the multiplicative complexity.
This is explained in part
due to the well-developed multiplicative complexity theory,
championed by Winograd~\cite{winograd1980complexity} and Heideman~\cite{heideman1899complexity}.
Such developments allowed the prediction of
theoretical lower bounds for the number of multiplications
required by some discrete transforms~\cite{heideman1899complexity,feig1992complexity}.
Much of the research in this field
is concerned the design of algorithms that
can be considered ``optimal''
in a multiplicative complexity measurement.

However,
bounded by theoretical constraints,
exact methods could only achieve
the prognosticated complexity minima
at best.
Even in such an optimal scenario,
floating-point operations are
involved.
In particular,
floating-point multiplications
are known to
possess relatively slow implementations,
even in hardware~\cite{liang2001fast}.

One way to circumvent a possibly significant multiplicative complexity
is to consider, not exact, but approximate computations.
In this case,
the
theoretical limits on the multiplicative complexity
do not apply.
A trade-off between complexity and accuracy may take place.

Several approximation methods have been proposed in literature.
Arithmetic transform procedures
explore nonuniform sampling and
number-theoretic functions to devise
multiplication-free algorithms for the DFT~\cite{reed1992simplified}, the DHT~\cite{cintra2002interpolate},
and the DCT~\cite{cintra2010act}.
Approximations for the DHT
based on Ramanujan numbers~\cite{bhatnagar1997friendly}
and
on wavelets~\cite{dee2001approximate}
were also suggested.
The DCT computation was shown to be approximated
in many ways.
In particular,
integer approximations
are a significant category of methods~\cite{merhav1999approximate,lengwehasatit1998variable,%
hossen1997approximate,natarajan1995approximate,chan2006minimal,malvar2003low,wahid2005h264}.
In~\cite{oraintara2002integer},
the DFT was submitted to an integer approximation study as well.

Integer approximation procedures constitute a class of practical interest.
Essentially,
these methods take advantage of
the fast computation of the integer arithmetic,
when compared to floating-point manipulations.
With the use
of dyadic rational approximations~\cite{britanak2007discrete}
and
the canonical signed digit representation~\cite{xu2007hamming},
integer multiplications
can be
elegantly converted
into combinations of
additions and bit shifting operations.
As a consequence,
multiplication counts are virtually zeroed
and,
in its place,
the number of additions and shifts are often quantified.

Another aspect of this discussion
concerns
usual requirements of orthogonality and perfect reconstruction.
Much emphasis has been put in these properties,
which frequently impose challenging
design constraints
for
integer approximation algorithms~\cite{britanak2007discrete}.
In particular,
orthogonal integer approximations for large blocklengths
can be difficult to be obtained.
Several existing design procedures require
the solution of large constrained non-linear
optimization problems in integer domain.
Even for small blocklengths
and considering exhaustive search,
solutions are not trivial~\cite{britanak2007discrete,oraintara2002integer}.

On the other hand,
nonorthogonal methods are becoming increasingly popular~\cite{koyuturk2006nonorthogonal}.
Classes of nonorthogonal transforms have been defined~\cite{french1974nonorthogonal}
and
algorithms for designing nonorthogonal basis have been considered~\cite{yu1992nonorthogonal}.
Nonorthogonal, but closely orthogonal, matrices
have found applications in soft clustering analysis~\cite{ding2005equivalence}.
Recently, blind source separation procedures were given a comprehensive treatment
with nonorthogonal matrices~\cite{fadaili2007nonorthogonal}.
Nonorthogonal basis images were also explored as a means to provide better representation methods
for compressed images~\cite{mikhael1996nonorthogonal}.

In this context,
the goal of the present work is the proposal of an
integer approximation method for discrete sinusoidal transforms
based on dyadic rational approximations.
In this study,
we initially relax,
but not neglect,
orthogonality and perfect reconstruction
constraints.
Subsequently,
we submit the proposed approximate transforms to a convenient
orthogonalization method
based on the matrix polar decomposition~\cite{higham1986applications}.
Related fast algorithms are suggested.
We also aim at
introducing
new square wave transforms
in a comparable fashion as studied
in~\cite{britanak2007discrete,haweel2001square,bouguezel2008low,lengwehasatit2004scalable}.

The paper is organized as follows.
In Section~\ref{sec.approximate},
a dyadic rational approximation
of the cosine function is examined.
Afterwards,
integer approximations for the
DFT, DHT, and DCT matrices are proposed;
and an optimized global scaling factor is considered.
In Section~\ref{sec.inverse},
the inverse transformation and
orthogonality issues are discussed;
an error analysis is also derived.
Section~\ref{sec.applications}
suggests
some potential applications for the proposed approximations.
Finally,
Section~\ref{sec.conclusions}
concludes the paper.

\section{Approximating Procedure for Discrete Sinusoidal Transforms}
\label{sec.approximate}

\subsection{Dyadic rational approximations}

The nearest integer function offers a possible venue to map
the values of the
transformation kernel into dyadic rational numbers.
This function simply returns integer values
according to the following construction:
\begin{align}
[x]
\triangleq
\mathrm{sgn}(x)
\left\lfloor
|x|+\frac{1}{2}
\right\rfloor,
\end{align}
where $\lfloor\cdot\rfloor$ is the floor function,
$\mathrm{sgn}(\cdot)$ is the sign function,
and
$|\cdot|$ returns the absolute value.
This definition
is in agreement to the implementation of the
rounding algorithm
available in the
standard mathematical library of C language.
A complex number $z = x+jy$, where $x,y\in \mathbb{R}$,
is rounded off according to
$[z]
\triangleq
[x]
+
j
[y]$.
More generally,
let the $m$th order dyadic rational approximating function
be defined as
\begin{align}
[x]_m
\triangleq
\frac{[2^m x]}{2^m},
\end{align}
where $x$ is a real number and $m$ is a nonnegative integer.
When $m=0$, the approximating function $[\cdot]_0$ is
equal to the nearest integer function.
Intuitively,
as $m\to\infty$,
we have that $[\cdot]_\infty$ becomes
the identity function.

Thus,
the $m$th order dyadic rational approximation of
a given kernel function can be obtained according to
\begin{align}
\kernel_m(i,k,n)
\triangleq
[\ker(i,k,n)]_m.
\end{align}
As a consequence,
transform vector $\mathbf{v}'$
can be approximated as $\hat{\mathbf{v}}'$,
whose components are expressed as follows:
\begin{equation}
\label{forward_rkernel}
\hat{v}_k' = \sum_{i=0}^{n-1}
v_i
\cdot
\kernel_m\left( i,k,n \right),
\quad
k=0,1,\ldots, n-1.
\end{equation}

For $m=0$,
the approximation is derived via
the usual rounding off operation.
In this case,
the values of $\kernel_0(\cdot,\cdot,\cdot)$
are $0$ or $\pm1$.
Although this mapping can be regarded as a coarse approximation for
$\kernel(\cdot,\cdot,\cdot)$,
it has zero multiplicative complexity,
and only addition operations are necessary
to render
the approximate transformed signal.
Another particularly interesting case occurs when $m=1$.
In this situation,
$\kernel_1(\cdot,\cdot,\cdot)$ returns only
$0$, $\pm1$, or $\pm2^{-1}$.
Thus, only additions and simple bitwise shift operations
are employed to obtain the approximate transform.
For higher values of $m$,
canonical signed digit representation could be applied
to furnish
multiplierless computations for
the quantities displayed in
Equation~(\ref{forward_rkernel}).
Clearly,
as $m\to\infty$,
$\kernel_m(\cdot,\cdot,\cdot) \to \kernel(\cdot,\cdot,\cdot)$,
where the limiting process indicates pointwise convergence~\cite{krantz2005analysis}.

Obviously,
different functions can possess the same
nearest integer approximation.
Thus
we may examine how
good
the approximations given by the
function $[\cdot]_m$ are.
Due to its ubiquity in discrete transform theory,
let us consider the cosine function to be approximated.
A multiplicative scaling factor $\alpha$
is also included for
an additional degree of freedom.
Adopting the mean square error
as the objective function to be minimized,
we can set up the following unconstrained optimization
problem:
\begin{align}
\min_\alpha
\int_{-\pi}^\pi
\left(
\alpha\cos(x)
-
[\alpha\cos(x)]_m
\right)^2
\mathrm{d}x,
\end{align}
for a fixed $m$.

Routine manipulations,
which employ standard optimization methods,
show that the optimal scaling factor
is a root of the following
non-linear equation:
\begin{align}
\label{pi.formula}
\frac{\pi}{2} \alpha^2
=
\frac{1}{4^m}
\sum_{k=0}^{2^m -1}
\sqrt{4^{m+1}{\alpha}^2 - (2k+1)^2}.
\end{align}
For $m=0$,
we have that the optimal scaling factor
is exactly
$\frac{\sqrt{2}}{\pi}\sqrt{4+\sqrt{16-\pi^2}}\approx1.1455$.
Since,
for larger values of $m$,
analytical computations  are beyond purpose,
Table~\ref{table.alpha.optimal}
lists optimal values obtained by numerical methods.
When $m\to\infty$,
the optimal scaling factor collapses to 1,
as expected.
As a byproduct, in this case,
Equation~(\ref{pi.formula}) furnishes
an infinite summation formula for the value of $\pi$
when $\alpha=1$.

\begin{table}
\centering
\caption{Optimal scaling}
\label{table.alpha.optimal}
\begin{tabular}{cc}
\noalign{\smallskip}
\hline
\noalign{\smallskip}
$m$      &  $\alpha$ \\
\noalign{\smallskip}
\hline
\noalign{\smallskip}
$0$      &  $1.1455$ \\
$1$      &  $1.0754$ \\
$2$      &  $1.0385$ \\
$3$      &  $1.0196$ \\
$4$      &  $1.0098$ \\
$\infty$ &  1        \\
\noalign{\smallskip}
\hline
\end{tabular}
\end{table}

\begin{figure*}
\centering
\subfigure[]{\epsfig{file=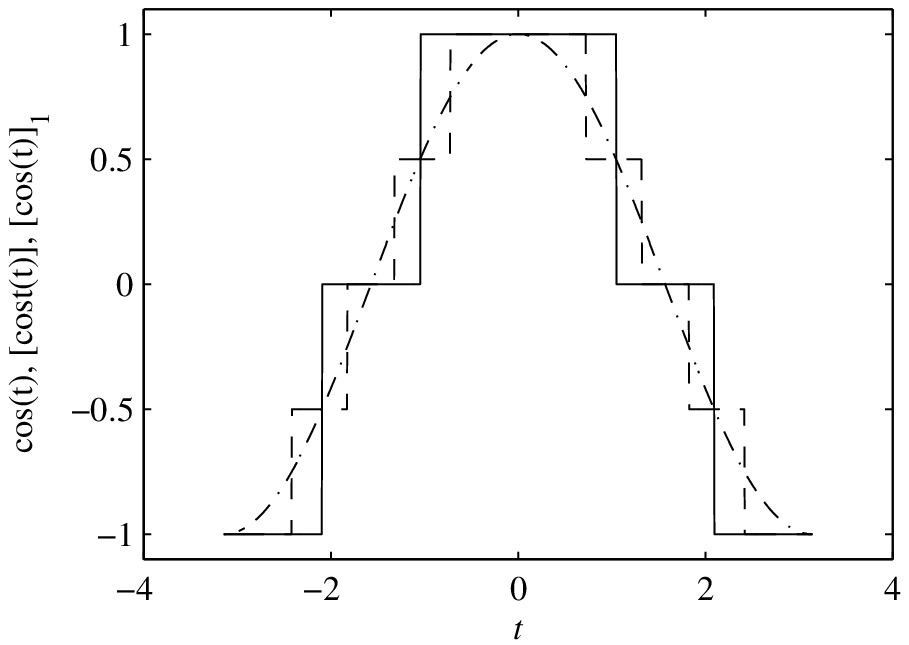}}
\subfigure[]{\epsfig{file=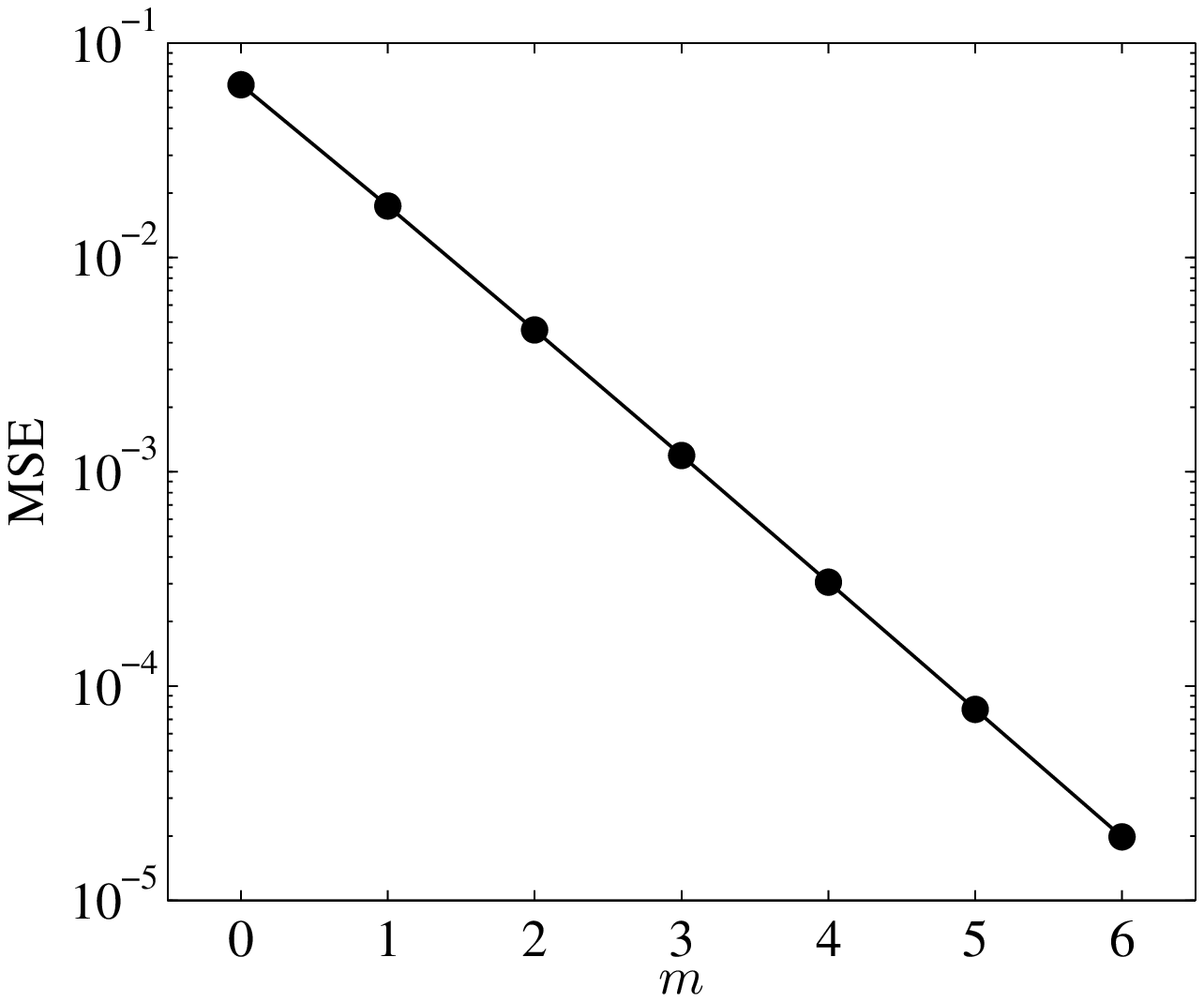,height=0.3672\linewidth}}
\caption{%
(a) Cosine function (dash-dotted line) and two approximations: $[\cos(t)]$ (solid line) and $[\cos(t)]_1$ (dashed line),
over the interval $-\pi<t\leq \pi$.
(b) Mean square error of $[\cos(t)]_m$ with respect to $\cos(t)$
in the interval $-\pi<t\leq\pi$ for several values of $m$.}
\label{fig.mse.cosine}
\end{figure*}

Despite of optimality issues,
for several practical short blocklengths and small approximation orders,
we have that
$\ker_m(\cdot,\cdot,n) = [\alpha \cdot \ker(\cdot,\cdot,n)]_m$,
for an optimal value $\alpha$.
Thus,
for operational purposes,
we could admit $\alpha=1$.
Figure~\ref{fig.mse.cosine}(a)
displays the plots of $[\cos(t)]_m$,
for $m = 0, 1$,
compared to $\cos(t)$,
over the interval $[-\pi, \pi]$.

In this case,
the mean square error
due to approximating
the cosine function
as
$[\cos(t)]_m$,
$-\pi<t\leq\pi$,
has
a closed formula given by:
\begin{align}
\begin{split}
\mse([\cos(t)]_m,\cos(t))
=&
\frac{\pi}{4}
-
\frac{1}{4^m}
\sum_{k=0}^{2^m-1}
\sqrt{4^{m+1}-(2k+1)^2}
\\
&
+
\frac{1}{4^m}
\sum_{k=0}^{2^m-1}
(2k+1)
\cos^{-1}
\left(
\frac{2k+1}{2^{m+1}}
\right).
\end{split}
\end{align}
Figure~\ref{fig.mse.cosine}(b) depicts the above mean square error
calculated
for $0\leq m\leq 6$.

Henceforth,
we adopt matrix notation.
Thus,
let the Fourier, Hartley, and cosine transformation matrices
be defined,
in their unitary form,
as
\begin{gather}
\mathbf{F}_n
=
\frac{1}{\sqrt{n}}
\left\{
\exp\left(-\frac{2\pi i jk}{n}\right)
\right\}_{(i,k)=0}^{n-1},
\\
\mathbf{H}_n
=
\frac{1}{\sqrt{n}}
\left\{
\cas\left(\frac{2\pi ik}{n}\right)
\right\}_{(i,k)=0}^{n-1},
\\
\mathbf{C}_n
=
\sqrt{\frac{2}{n}}
\left\{
\left(1 - (1-1/\sqrt{2})\delta_{0,k} \right)
\cos\left(\frac{\pi (k+1/2)i}{n}\right)
\right\}_{(i,k)=0}^{n-1},
\end{gather}
respectively.

In account of the previous discussion,
we define the
$m$th order dyadic rational matrices associated to
$\mathbf{F}_n $,
$\mathbf{H}_n $,
and
$\mathbf{C}_n $
according to
\begin{align}
\mathbf{F}_n^{(m)}
&\triangleq
\left[
\sqrt{n}
\mathbf{F}_n
\right]_m,
\\
\mathbf{H}_n^{(m)}
&\triangleq
\left[
\sqrt{\frac{n}{2}}
\mathbf{H}_n
\right]_m,
\\
\mathbf{C}_n^{(m)}
&\triangleq
\left[
\sqrt{\frac{n}{2}}
\mathbf{C}_n
\right]_m,
\end{align}
where the operator $[\cdot]_m$ when applied to matrices acts componentwisely.
Prior to the application of the dyadic rational approximation procedure,
the elements of original transform matrices were
subject to a normalization by
$\sqrt{n}$ or $\sqrt{n/2}$.
Moreover,
when scaled by $2^m$,
the resulting matrices are constituted
of integer numbers only.
For instance,
when $m=0$,
each entry of the dyadic rational matrices is $-1$, $0$, or $+1$.

\subsection{Optimization}

Indeed,
the proposed dyadic rational matrices can furnish approximations
for
$\mathbf{F}_n$,
$\mathbf{H}_n$,
and
$\mathbf{C}_n$.
A possible way
to obtain
such approximations
is by the inclusion of a scaling factor,
as suggested in~\cite[p.~275]{britanak2007discrete}.
Scaling factors are introduced
in such a way
to minimize
a chosen
error measure
between
an original transformation matrix
and
its approximation.

In matrix terms,
the dyadic rational matrices link a vector $\mathbf{v}$
to its approximate transform $\hat{\mathbf{v}}'$,
according to
the following expression
\begin{align}
\hat{\mathbf{v}}'
=
\beta
\cdot
\mathbf{K}_n^{(m)} \mathbf{v},
\end{align}
where
$\mathbf{K}_n^{(m)} \in \{ \mathbf{F}_n^{(m)}, \mathbf{H}_n^{(m)}, \mathbf{C}_n^{(m)} \}$,
for a fixed $m$,
and
$\beta$ is a real number.
The quantity $\beta$ provides a global adjustment in such a way that
$\beta\mathbf{K}_n^{(m)}$
satisfactorily
approximates
$\mathbf{K}_n \in \{ \mathbf{F}_n, \mathbf{H}_n, \mathbf{C}_n \}$,
respectively.

Thus,
we have set the following unconstrained optimization problem:
\begin{align}
\min_\beta
\mathrm{Error}
\left(\mathbf{K}_n, \beta \mathbf{K}_n^{(m)} \right),
\end{align}
where
$\mathrm{Error}(\cdot, \cdot)$
quantifies the dissimilarity
between its arguments,
according to a selected measure.
Adopting the Frobenius norm of the matrix difference
$\mathbf{K}_n - \beta \mathbf{K}_n^{(m)}$
as the objective function to be minimized~\cite[p.~523]{seber2008handbook},
we obtain optimal values for the scaling factor $\beta$ using
conventional optimization procedures.
Tables~\ref{table.beta.optimal.dft},
\ref{table.beta.optimal.dht},
and
\ref{table.beta.optimal.dct}
show the numerically computed optimal values of $\beta$ for
selected practical blocklengths
at various approximation orders.

\begin{table*}
\centering
\caption{Optimal $\beta$ for the approximate DFT}
\label{table.beta.optimal.dft}
\begin{tabular}{ccccccc}
\noalign{\smallskip}
\hline
\noalign{\smallskip}
      & \multicolumn{6}{c}{$m$}\\
\cline{2-7}
\noalign{\smallskip}
$n$   & $0$      & $1$      & $2$      & $3$      & $4$      & $\infty$\\
\noalign{\smallskip}
\hline
\noalign{\smallskip}
$4$   & $0.5000$ & $0.5000$ & $0.5000$ & $0.5000$ & $0.5000$ & $1/2$ \\
$6$   & $0.3513$ & $0.3864$ & $0.4255$ & $0.4068$ & $0.4068$ & $1/\sqrt{6}$ \\
$8$   & $0.3121$ & $0.3745$ & $0.3480$ & $0.3480$ & $0.3560$ & $1/\sqrt{8}$ \\
$12$  & $0.2478$ & $0.2732$ & $0.3008$ & $0.2877$ & $0.2877$ & $1/\sqrt{12}$ \\
$16$  & $0.2169$ & $0.2548$ & $0.2386$ & $0.2488$ & $0.2511$ & $1/4$ \\
$24$  & $0.1706$ & $0.1977$ & $0.2076$ & $0.2011$ & $0.2050$ & $1/\sqrt{24}$ \\
$32$  & $0.1488$ & $0.1741$ & $0.1704$ & $0.1757$ & $0.1771$ & $1/\sqrt{32}$ \\
$64$  & $0.1046$ & $0.1195$ & $0.1202$ & $0.1242$ & $0.1252$ & $1/8$ \\
$128$ & $0.0734$ & $0.0838$ & $0.0854$ & $0.0879$ & $0.0883$ & $1/\sqrt{128}$ \\
\noalign{\smallskip}
\hline
\end{tabular}
\end{table*}

\begin{table*}
\centering
\caption{Optimal $\beta$ for the approximate DHT}
\label{table.beta.optimal.dht}
\begin{tabular}{ccccccc}
\noalign{\smallskip}
\hline
\noalign{\smallskip}
      & \multicolumn{6}{c}{$m$}\\
\cline{2-7}
\noalign{\smallskip}
$n$   & $0$      & $1$      & $2$      & $3$      & $4$      & $\infty$\\
\noalign{\smallskip}
\hline
\noalign{\smallskip}
$4$   & $0.5000$ & $1.0000$ & $0.6666$ & $0.6666$ & $0.7273$ & $1/\sqrt{2}$ \\
$6$   & $0.4509$ & $0.6095$ & $0.5511$ & $0.5511$ & $0.5944$ & $1/\sqrt{3}$ \\
$8$   & $0.3745$ & $0.6243$ & $0.4780$ & $0.4779$ & $0.5105$ & $1/2$ \\
$12$  & $0.3189$ & $0.4310$ & $0.3897$ & $0.3897$ & $0.4203$ & $1/\sqrt{6}$ \\
$16$  & $0.2800$ & $0.3839$ & $0.3328$ & $0.3465$ & $0.3575$ & $1/\sqrt{8}$ \\
$24$  & $0.2300$ & $0.2950$ & $0.2811$ & $0.2784$ & $0.2947$ & $1/\sqrt{12}$ \\
$32$  & $0.2011$ & $0.2529$ & $0.2392$ & $0.2466$ & $0.2513$ & $1/4$ \\
$64$  & $0.1445$ & $0.1709$ & $0.1694$ & $0.1749$ & $0.1774$ & $1/\sqrt{32}$ \\
$128$ & $0.1026$ & $0.1192$ & $0.1206$ & $0.1240$ & $0.1250$ & $1/8$ \\
\noalign{\smallskip}
\hline
\end{tabular}
\end{table*}

\begin{table*}
\centering
\caption{Optimal $\beta$ for the approximate DCT}
\label{table.beta.optimal.dct}
\begin{tabular}{ccccccc}
\noalign{\smallskip}
\hline
\noalign{\smallskip}
      & \multicolumn{6}{c}{$m$}\\
\cline{2-7}
\noalign{\smallskip}
$n$   & $0$      & $1$      & $2$      & $3$      & $4$      & $\infty$\\
\noalign{\smallskip}
\hline
\noalign{\smallskip}
$4$   & $0.5511$ & $0.7363$ & $0.6478$ & $0.7006$ & $0.7133$ & $1/\sqrt{2}$ \\
$6$   & $0.4462$ & $0.5955$ & $0.5682$ & $0.5572$ & $0.5873$ & $1/\sqrt{3}$ \\
$8$   & $0.3922$ & $0.4891$ & $0.4831$ & $0.4925$ & $0.5014$ & $1/2$        \\
$12$  & $0.3303$ & $0.3920$ & $0.3929$ & $0.4065$ & $0.4090$ & $1/\sqrt{6}$ \\
$16$  & $0.2876$ & $0.3320$ & $0.3380$ & $0.3497$ & $0.3548$ & $1/\sqrt{8}$ \\
$24$  & $0.2342$ & $0.2709$ & $0.2816$ & $0.2853$ & $0.2889$ & $1/\sqrt{12}$ \\
$32$  & $0.2038$ & $0.2366$ & $0.2424$ & $0.2481$ & $0.2495$ & $1/4$ \\
$64$  & $0.1456$ & $0.1657$ & $0.1728$ & $0.1750$ & $0.1762$ & $1/\sqrt{32}$ \\
$128$ & $0.1030$ & $0.1172$ & $0.1223$ & $0.1241$ & $0.1247$ & $1/8$ \\
\noalign{\smallskip}
\hline
\end{tabular}
\end{table*}

Other error measures,
such as the spectral norm or the $p$-norm for $p\neq2$,
could be considered instead of the Frobenius norm.
In that case,
different, but comparable, values of $\beta$ would be found.
Since $\beta$ is an overall scaling factor,
it does not affect further mathematical considerations
on the nature of the approximate matrices
$\mathbf{K}_n^{(m)}$.

\begin{figure*}
\centering
\subfigure[$m=\infty$]{\epsfig{file=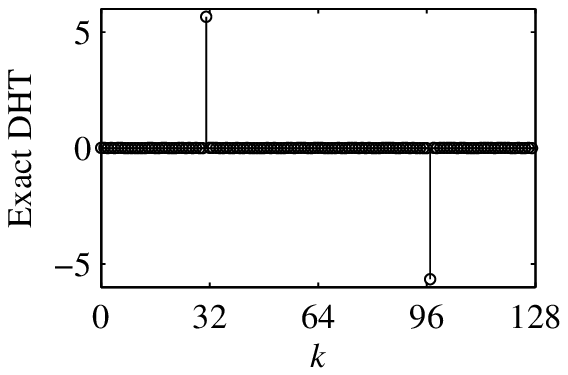,width=0.32\linewidth}}
\subfigure[$m=0$]{\epsfig{file=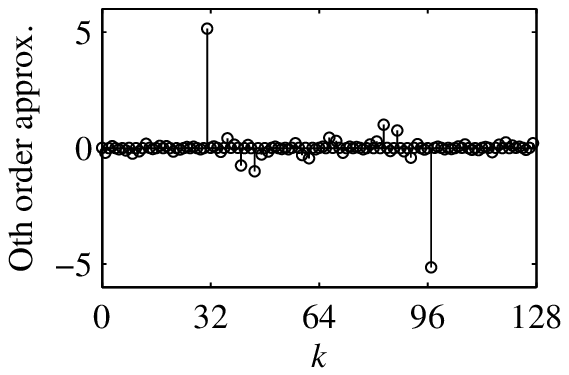,width=0.32\linewidth}}
\subfigure[$m=1$]{\epsfig{file=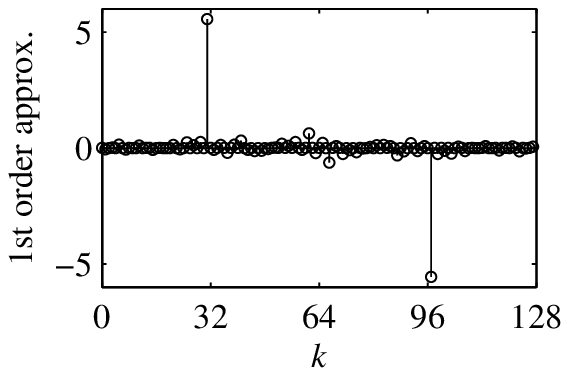,width=0.32\linewidth}}
\\
\subfigure[$m=2$]{\epsfig{file=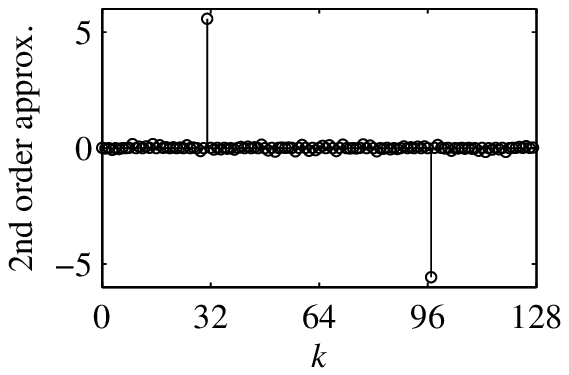,width=0.32\linewidth}}
\subfigure[$m=3$]{\epsfig{file=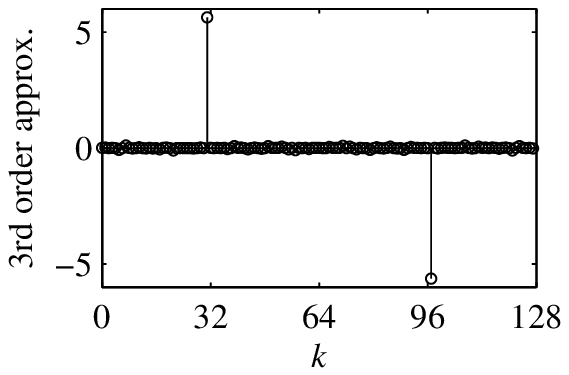,width=0.32\linewidth}}
\subfigure[$m=4$]{\epsfig{file=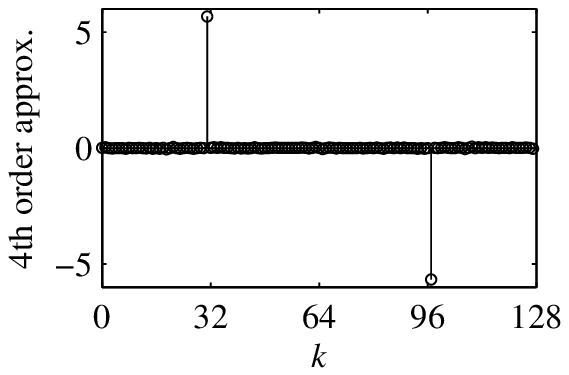,width=0.32\linewidth}}
\caption{Discrete Hartley transform of a pure sinusoidal signal:
(a) exact computation,
(b) 0th order approximation,
(c) 1st order approximation,
(d) 2nd order approximation,
(e) 3rd order approximation,
and
(f) 4th order approximation.}
\label{fig.test.rdht}
\end{figure*}

For illustrative purposes,
consider a pure sinusoidal signal $v_i=\sin\left(2\pi\frac{31}{128}i\right)$, $i=0,\ldots,127$.
Figure~\ref{fig.test.rdht}
shows
the DHT of $\mathbf{v}$,
compared to
the approximations
given by the discussed method for
$m=0,1,\ldots,4$.
As $m$ increases,
approximations become more accurate.
Approximations for the DFT and the DCT
showed similar behavior.

\section{Inverse Transformation}
\label{sec.inverse}

\subsection{Matrix invertibility and nonorthogonality}

Although in several applications,
such as pattern classification based on transform domain
feature extraction~\cite{jing2004palmprint}
and transform adaptive filtering~\cite[p.~154]{diniz1996adaptive},
only the forward transform is required,
we investigate the inverse transformation of the
proposed approximate method.

\begin{figure*}
\centering
\epsfig{file=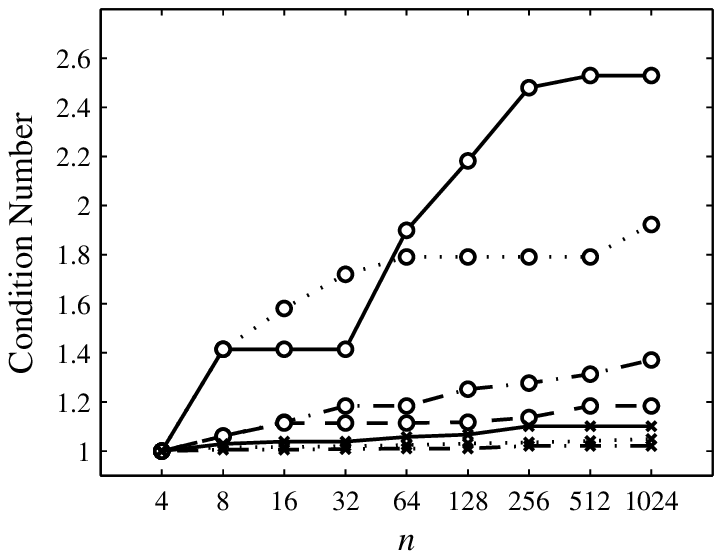}
\caption{%
Condition number of approximate Fourier matrices for
$m=0$ (--$\circ$--),
$m=1$ ($\cdots\circ\cdots$),
$m=2$ ($\cdot-\!\!\circ\!\!-\cdot$),
$m=3$ (-\,-$\circ$-\,-),
$m=4$ (--$\times$--),
$m=5$ ($\cdots\!\!\!\times\!\!\!\cdots$),
and
$m=6$ ($\cdot-\!\!\!\!\times\!\!\!\!-\cdot$).
}
\label{fig.cond.num.fourier}
\end{figure*}

Function $[\cdot]_m$
imposes inherent analytical difficulties.
Consequently,
it may be not obvious to establish whether,
for any given value of $n$ and $m$,
the inverse matrix $\left(\mathbf{K}_n^{(m)}\right)^{-1}$ does exist.
Then, we resort to exhaustive computational search.
For $n\leq1024$ and $m\leq6$,
the inverse of
$\mathbf{K}_n^{(m)}$
was always found to exist.
The evaluation of the condition number is adopted as a means
to assess how well-conditioned these matrices are
in terms of matrix inversion~\cite{ducroz1992stability}.
Indeed,
the condition number measures
the sensitivity of matrix inversion~\cite{higham1995condition}.
Thus,
for the considered search space,
the 2-norm condition number of
$\mathbf{K}_n^{(m)}$
is small,
never exceeding
$2.5295$, $2.5295$, and $2.9432$,
for the
Fourier, Hartley, and cosine approximate matrices,
respectively.
Such low values of the condition number indicate well-conditioned matrices.
Figure~\ref{fig.cond.num.fourier}
depicts the values of the condition number
of $\mathbf{F}_n^{(m)}$
as a function of the transform size $n$.
Approximate matrices associated to the Fourier and Hartley transforms
had identical condition number
and matrices $\mathbf{C}_n^{(m)}$ presented similar values.
For comparison,
notice that since the exact Fourier, Hartley, and cosine
matrices are unitary,
their condition numbers are equal to one
for any blocklength.

Thus,
for practical purposes,
assuming that
the inverse of
$\mathbf{K}_n^{(m)}$
is well-defined,
the following manipulation holds true:
\begin{align}
\label{eq.holds.true}
(\mathbf{K}_n^{(m)})^H
\left(
\mathbf{K}_n^{(m)}
(\mathbf{K}_n^{(m)})^H
\right)^{-1}
\mathbf{K}_n^{(m)}
=
\mathbf{I}_n,
\end{align}
where the superscript $H$ indicates the Hermitian transposition
and
$\mathbf{I}_n$ is the identity matrix of size~$n$.
Consequently,
we conclude that
\begin{align}
\left(
\mathbf{K}_n^{(m)}
\right)^{-1}
=
(\mathbf{K}_n^{(m)})^H
\left(
\mathbf{K}_n^{(m)}
(\mathbf{K}_n^{(m)})^H
\right)^{-1}
.
\end{align}
Strictly,
the proposed approximate matrices lack
unitary property,
since
$\left(\mathbf{K}_n^{(m)}\right)^{-1}\not=(\mathbf{K}_n^{(m)})^H$.
An extra multiplicative term
$\mathbf{D}^{-1} \triangleq  \left( \mathbf{K}_n^{(m)}(\mathbf{K}_n^{(m)})^H \right)^{-1}$
is necessary
to furnish the matrix inversion,
which enables perfect signal reconstruction.
We have then obtained the following set of relations:
\begin{align}
\label{eq.method.1.fwd}
\hat{\mathbf{v}}' &= \beta \cdot \mathbf{K}_n^{(m)} \mathbf{v}, \\
\label{eq.method.1.inv}
\mathbf{v} &= \frac{1}{\beta} \cdot  (\mathbf{K}_n^{(m)})^H
\mathbf{D}^{-1}
\hat{\mathbf{v}}'.
\end{align}

Except for the presence of $\mathbf{D}^{-1}$,
both forward and inverse approximate transformations
share the same computational complexity.
This is because
their matrices are related by a simple transposition.

However,
if $m\to\infty$,
then
$\mathbf{K}_n^{(m)}$
converges to
$\mathbf{K}_n\in\{ \mathbf{F}_n, \mathbf{H}_n, \mathbf{C}_n \}$.
Being $\mathbf{K}_n$ a unitary matrix,
we obtain that
\begin{align}
\lim_{m\to\infty}
\mathbf{K}_n^{(m)}
(\mathbf{K}_n^{(m)})^H
=
\mathbf{K}_n
\mathbf{K}_n^H
=
\mathbf{I}_n.
\end{align}
Thus,
$\mathbf{K}_n^{(m)}$ is asymptotically unitary in terms of $m$.
Therefore,
for sufficiently large $m$,
Equations~(\ref{eq.method.1.fwd}) and~(\ref{eq.method.1.inv})
can be recast as:
\begin{align}
\hat{\mathbf{v}}' &= \beta \cdot \mathbf{K}_n^{(m)} \mathbf{v}, \\
\hat{\mathbf{v}} & \approx \mathbf{v} = \frac{1}{\beta} \cdot (\mathbf{K}_n^{(m)})^H \hat{\mathbf{v}}',
\end{align}
where $\hat{\mathbf{v}}$ is an approximation due to the nonorthogonality
of $\mathbf{K}_n^{(m)}$.

\subsection{Orthogonalization}

In view of the above discussion,
we can re-examine the optimization
procedure performed in the previous section.
There,
the approximate matrices were
corrected by an overall constant factor termed $\beta$.
In contrast to that,
we can consider a more refined correction term.
Notice that the term
$\mathbf{K}_n^{(m)}(\mathbf{K}_n^{(m)})^H$
is a Gram matrix,
which is Hermitian,
non-negative definite.
Then,
$\mathbf{D}^{-1}$ is also Hermitian~\cite[p.~82]{seber2008handbook}
and
its matrix square root
is well-defined, Hermitian~\cite[p.~82]{seber2008handbook},
and unique~\cite[p.~89]{johnson1990matrix}.
Consequently,
Equation~(\ref{eq.holds.true})
becomes:
\begin{align}
(\mathbf{K}_n^{(m)})^H
\left(
\mathbf{D}^{-1}
\right)^{\frac{1}{2}}
\left(
\mathbf{D}^{-1}
\right)^{\frac{1}{2}}
\mathbf{K}_n^{(m)}
=
\mathbf{I}_n.
\end{align}
Taking into account that
the inverse of a Hermitian matrix is also Hermitian~\cite[p.~82]{seber2008handbook},
we have that
\begin{align}
\left(
\left(
\mathbf{D}^{-1}
\right)^{\frac{1}{2}}
\mathbf{K}_n^{(m)}
\right)^H
\left(
\mathbf{D}^{-1}
\right)^{\frac{1}{2}}
\mathbf{K}_n^{(m)}
=
\mathbf{I}_n.
\end{align}
It follows that
$\left( \mathbf{D}^{-1} \right)^{\frac{1}{2}} \mathbf{K}_n^{(m)}$
is a unitary matrix;
and the term
$\mathbf{S}^{-1} \triangleq \left( \mathbf{D}^{-1} \right)^{\frac{1}{2}}$
adjusts the approximate matrix.
In fact,
this adjustment term is exactly the inverse of the unique Hermitian non-negative definite matrix
obtained when
$\mathbf{K}_n^{(m)}$
is factorized according to the
matrix polar decomposition procedure~\cite[p.~348]{seber2008handbook}.

Being the unique unitary matrix of a polar decomposition,
the term
$\mathbf{S}^{-1}\mathbf{K}_n^{(m)}$
possesses some optimal approximation properties.
In particular,
when the Frobenius norm is utilized as a distance measure,
$\mathbf{S}^{-1}\,\mathbf{K}_n^{(m)}$
is the nearest unitary matrix to
$\mathbf{K}_n^{(m)}$~\cite{higham1986applications}.

Thus,
Equations~(\ref{eq.method.1.fwd}) and~(\ref{eq.method.1.inv})
can be modified into the following form:
\begin{align}
\label{eq.method.2.fwd}
\hat{\mathbf{v}}' &= \mathbf{S}^{-1} \,\mathbf{K}_n^{(m)} \mathbf{v}, \\
\mathbf{v} &=  (\mathbf{K}_n^{(m)})^H \mathbf{S}^{-1} \hat{\mathbf{v}}', \nonumber
\end{align}
where the scaling factor $\beta$
was suppressed and its role is taken by $\mathbf{S}^{-1}$.

Regarding the closeness of these approximations to the exact matrices,
computational calculations for $n\leq1024$ and $m = 0,1,\ldots, 4$
confirm that
\begin{align}
\|\mathbf{K}_n -  \mathbf{S}^{-1}\, \mathbf{K}_n^{(m)}   \|_F
<
\|\mathbf{K}_n -  \beta \mathbf{K}_n^{(m)}   \|_F,
\end{align}
where $\|\cdot\|_F$ denotes the Frobenius norm.
In a sense,
this could be expected,
since the adjustment offered by the optimal scaling factor $\beta$
corresponds simply to the matrix
$\beta \cdot \mathbf{I}_n$.
On the other hand,
the adjustment by
$\mathbf{S}^{-1}$ possesses a higher arithmetic complexity,
which allows a better approximation.

Therefore,
a significant part of
the computational cost of
using
$\mathbf{S}^{-1} \mathbf{K}_n^{(m)}$
as an approximation
for
$\mathbf{K}_n$
relies on the
complexity of
$\mathbf{S}^{-1}$.
This observation prompts us to examine
the behavior of $\mathbf{S}^{-1}$.

\subsection{Error analysis}

Being $\mathbf{K}_n$
already unitary matrices,
when
$\eta\mathbf{K}_n$ are
submitted to a polar decomposition,
one obtains that
\begin{align}
\eta \cdot \mathbf{K}_n =  (\eta\cdot\mathbf{I}_n) \cdot \mathbf{K}_n,
\end{align}
where $\eta$ is a normalizing factor equal
to $\sqrt{n}$ for the DFT
and
to $\sqrt{n/2}$ for the DHT or DCT.
Introducing a perturbation matrix
$\Delta\mathbf{E}_1$,
we can represent the discussed dyadic approximation
according to
$\mathbf{K}_n^{(m)} = \eta\cdot\mathbf{K}_n + \Delta\mathbf{E}_1$.
Therefore,
we have that
polar decomposition of $\mathbf{K}_n^{(m)}$
is given by
\begin{align}
\eta\cdot\mathbf{K}_n + \Delta\mathbf{E}_1
=
\eta
\cdot
(
\mathbf{I}_n + \Delta\mathbf{S}
)
\cdot
(
\mathbf{K}_n + \Delta\mathbf{E}_2
),
\end{align}
where
$\Delta\mathbf{S}$
and
$\Delta\mathbf{E}_2$
are
induced
perturbation matrices in the Hermitian and orthogonal polar decomposition factors,
respectively.
The Hermitian matrix
is related to the adjustment matrix
as
$\mathbf{S} = \eta \cdot (\mathbf{I}_n + \Delta\mathbf{S})$.
The matrix
$\mathbf{K}_n + \Delta\mathbf{E}_2$
is recognized as an orthogonal approximation to
$\mathbf{K}_n$~\cite{higham1986applications,seber2008handbook}.
More explicitly,
we have that
\begin{align}
\mathbf{K}_n + \Delta\mathbf{E}_2
=&
\mathbf{S}^{-1}
\cdot
\mathbf{K}_n^{(m)}
\\
=&
\eta^{-1} \cdot (\mathbf{I}_n + \Delta\mathbf{S})^{-1}
\cdot
(\eta\mathbf{K}_n + \Delta\mathbf{E}_1)
,
\end{align}
where
$\Delta\mathbf{E}_2$ has minimum Frobenius norm~\cite{higham1986applications}.
Thus,
if
$\Delta\mathbf{S}$ approaches a null matrix,
then
$\eta\cdot\mathbf{S}^{-1}$ is close to an identity matrix.
This is desirable,
since
the complexity of an identity matrix is null.

Aiming to determine an upper bound for the distance between
$\eta\cdot\mathbf{S}^{-1}$ and $\mathbf{I}_n$,
we first analyze
the distance between its inverse
$\eta^{-1}\cdot\mathbf{S}$ and $\mathbf{I}_n$.
This latter distance is quantified by
the following approximation error
\begin{align}
\epsilon
&=
\frac{\| \eta^{-1}\mathbf{S} - \mathbf{I}_n \|_F}{\| \mathbf{I}_n \|_F}
\\
&=
\frac{\| \Delta\mathbf{S} \|_F}{\| \mathbf{I}_n \|_F}.
\end{align}
In a similar manner,
it is reasonable to quantify
the perturbation error induced by
$\Delta\mathbf{E}_1$
as
\begin{align}
\epsilon_1
=
\frac{\| \Delta\mathbf{E}_1 \|_F}{\| \eta\mathbf{K}_n \|_F}.
\end{align}
By construction,
the elements of $\Delta\mathbf{E}_1$
are bounded by $1/2^{m+1}$.
Thus,
in the worst possible scenario,
we have that
\begin{align}
\Delta\mathbf{E}_1
=
\frac{1}{2^{m+1}}
\mathbf{J}_n,
\end{align}
where
$\mathbf{J}_n$
is a square matrix of ones with dimension $n$.
This type of matrix and its properties are discussed
in~\cite[p.~2]{moser1996models}.
Therefore,
we obtain that
$\| \Delta\mathbf{E}_1 \|_F = \frac{n}{2^{m+1}}$.
Additionally,
for all considered discrete sinusoidal transforms,
we have that
$\|\mathbf{K}_n\|_F = \sqrt{n}$.
Thus,
we have that
\begin{align}
\epsilon_1
\leq
\frac{1}{2^{m+1}}
\frac{\sqrt{n}}{\eta}
=
\frac{1}{2^{m+1}}
\times
\left\{
\begin{array}{cl}
1, & \text{for the DFT,}
\\
\sqrt{2},
& \text{for the DHT or DCT}
\end{array}
\right\}
.
\end{align}

In~\cite{higham1986applications},
Higham submitted the polar decomposition procedure
to a comprehensive error
analysis.
It was demonstrated that error measures $\epsilon$ and $\epsilon_1$ could be related
according to
\begin{align}
\epsilon
\leq
\sqrt{2} \epsilon_1
+
O(\epsilon_1^2).
\end{align}
Performing the necessary substitutions and
observing that $\eta/\sqrt{n}$ is a constant,
we obtain:
\begin{align}
\epsilon
&
\leq
\frac{1}{2^{m}}
\times
\left\{
\begin{array}{cl}
1/\sqrt{2},& \text{for the DFT,}
\\
1,
& \text{for the DHT or DCT,}
\end{array}
\right\}
+
O
\left(
\frac{1}{4^m}
\right)
.
\end{align}
Therefore,
$\epsilon = O\left(\frac{1}{2^m}\right)$.

Now we return to the error analysis of $\eta\mathbf{S}^{-1}$.
Analogously,
the sought approximation error is given by
\begin{align}
\epsilon'
&=
\frac{\| \eta\mathbf{S}^{-1} - \mathbf{I}_n \|_F}{\| \mathbf{I}_n \|_F}.
\end{align}
Invoking results on the stability of matrix
inversion~\cite{golub1996matrix,ducroz1992stability},
we conclude that
$\epsilon'$ has the same asymptotic behavior as $\epsilon$.
Additionally,
the
asymptotic behaviors of
$\epsilon$ and $\epsilon'$ are independent
of the blocklength
and are related only to the approximation order $m$.
For a fixed $m$,
we have that
$\epsilon,\epsilon' = O(1)$.
On the other hand,
fixing the blocklength size,
we also obtain that
$\epsilon, \epsilon' = O(1/2^m)$.

\section{Applications}
\label{sec.applications}

\subsection{$8$-point approximate DFT}

Usual 8-point DFT is a fundamental building block of several
signal processing methods~\cite{lin2007processor,blahut1985fast}.
The  1st order $8$-point approximate DFT
possesses the following matrix transformation:
\begin{align}
\mathbf{F}_8^{(1)}
=
\frac{1}{2}
\left[\begin{smallmatrix}
   2   &          2      &  2         & 2        &  2   &    2     & 2        &    2      \\
   2   &          1 -  j &       -2j  &-1 - j    & -2   &  -1 + j  &       2j &    1 + j  \\
   2   &             -2j & -2         &       2j &  2   &     -2j  &-2        &    2j     \\
   2   &         -1 -1j  &        2j  & 1 - j    & -2   &   1 + j  &      -2j &   -1 + j  \\
   2   &         -2      &  2         &-2        &  2   &   -2     & 2        &   -2      \\
   2   &         -1 + 1j &       -2j  & 1 + j    & -2   &   1 - j  &       2j &   -1 - j  \\
   2   &              2j & -2         &      -2j &  2   &      2j  &-2        &    -2j     \\
   2   &          1 + 1j &        2j  &-1 + j    & -2   &  -1 - j  &      -2j &    1 - j
\end{smallmatrix}\right].
\end{align}

Applying usual methods for matrix factorization~\cite{blahut1985fast},
one can derive the following
construction:
\begin{align}
\mathbf{F}_8^{(1)}
=
\mathbf{P} \cdot \mathbf{A}_4 \cdot \mathbf{A}_3 \cdot \mathbf{T} \cdot \mathbf{A}_2 \cdot \mathbf{A}_1,
\end{align}
where each factor is defined according to
\begin{align}
\mathbf{A}_1 =
\left[\begin{smallmatrix}
1 &  &  &  &1 &  &  & \\
         &1 &  &  &  &1 &  & \\
         &  &1 &  &  &  &1 & \\
         &  &  &1 &  &  &  &1\\
       1 &  &  &  &- &  &  & \\
         &1 &  &  &  &- &  & \\
         &  &1 &  &  &  &- & \\
         &  &  &1 &  &  &  &-
\end{smallmatrix}\right],
\quad
\mathbf{A}_2 =
\left[\begin{smallmatrix}
1 &  &1 &  &  &  &  &  \\
         &1 &  &1 &  &  &  &  \\
       1 &  &- &  &  &  &  &  \\
         &1 &  &- &  &  &  &  \\
         &  &  &  &1 &  &  & \\
         &  &  &  &  &1 &  &1\\
         &  &  &  &  &  &1 & \\
         &  &  &  &  &1 &  &-
\end{smallmatrix}\right],
\end{align}
\begin{align}
\mathbf{A}_3 =
\left[\begin{smallmatrix}
1 & &  &  &  &  &  &  \\
         &1&  &  &  &  &   & \\
         & & 1&  &  &  &   & \\
         & &  & 1&  &  &   & \\
         & &  &  & 1&  & 1 & \\
         & &  &  &  & 1&   &1\\
         & &  &  & 1&  & - &  \\
         & &  &  &  & 1&   &-
\end{smallmatrix}\right],
\quad
\mathbf{A}_4 =
\left[\begin{smallmatrix}
1 &1 &  &  &  &  &  & \\
1 &- &  &  &  &  &  &  \\
  &  &1 &1 &  &  &  & \\
  &  &1 &- &  &  &  &  \\
  &  &  &  &1 &1 &  & \\
  &  &  &  &1 &- &  &  \\
  &  &  &  &  &  &1 &1\\
  &  &  &  &  &  &1 &-
\end{smallmatrix}\right],
\end{align}
\begin{align}
\mathbf{P} =
\left[\begin{smallmatrix}
    1 &  & &  &  &  &   & \\
      &  & &  & 1&  &   & \\
      &  &1&  &  &  &   & \\
      &  & &  &  &  & 1 & \\
      &1 & &  &  &  &   & \\
      &  & &  &  & 1&   & \\
      &  & & 1&  &  &   & \\
      &  & &  &  &  &   &1
\end{smallmatrix}\right],
\quad
\mathbf{T}
=
\left[\begin{smallmatrix}
    1 &  & &  &  &  &   & \\
      & 1& &  &  &  & \\
      &  &1&  &  &  &   & \\
      &  & &-j&  &  &   & \\
      &  & &  & 1&  &   & \\
      &  & &  &  & -j/2 &   & \\
      &  & &  &  &  & -j  & \\
      &  & &  &  &  &   &1/2
\end{smallmatrix}\right]
,
\end{align}
where ``$-$'' represents $-1$ and blank spaces are zeroes.
With this sparse matrix factorization,
the transformation has its computational complexity reduced to
26 additions, and only two bit shifts.

To illustrate one of the proposed approximation methods,
let us consider the nonorthogonal approach
as indicated in Equation~(\ref{eq.method.1.inv}).
The exact computation of the inverse transformation
$\mathbf{F}_8^{(1)}$
furnishes the following relation:
\begin{align}
(\mathbf{F}_8^{(1)})^{-1}
=
(\mathbf{F}_8^{(1)})^{H}
\mathbf{D}^{-1},
\end{align}
where
\begin{align}
16
\cdot
\mathbf{D}^{-1}
=&
\left[
\begin{smallmatrix}
     2  &     &      &      &      &      &      &     \\
        &  3  &      &      &      &  -   &      &     \\
        &     &   2  &      &      &      &      &     \\
        &     &      &   3  &      &      &      &  -  \\
        &     &      &      &   2  &      &      &     \\
        & -   &      &      &      &   3  &      &     \\
        &     &      &      &      &      &   2  &     \\
        &     &      &  -   &      &      &      &   3 \\
\end{smallmatrix}
\right]
=
2\cdot \mathbf{I}_8
+
\left[
\begin{smallmatrix}
 \phantom{1}  &   &             &   &             &   &             &     \\
              & 1 &             &   &             & - &             &     \\
              &   & \phantom{1} &   &             &   &             &     \\
              &   &             & 1 &             &   &             &  -  \\
              &   &             &   & \phantom{1} &   &             &     \\
              & - &             &   &             & 1 &             &     \\
              &   &             &   &             &   & \phantom{1} &     \\
              &   &             & - &             &   &             &   1 \\
\end{smallmatrix}
\right]
\end{align}
Notice the extremely simple expression for $\mathbf{D}^{-1}$
as well as its low computational complexity,
which requires only additions and bit shifts.

Surprisingly,
the inverse matrix
$\mathbf{F}_8^{(1)}$
is related to the zeroth order approximate
matrix
$\mathbf{F}_8^{(0)}$:
\begin{align}
(\mathbf{F}_8^{(1)})^{-1}
=
\frac{1}{8}
(\mathbf{F}_8^{(0)})^{H}.
\end{align}
\emph{Per se},
this relation constitutes
the basis for
a new complex-valued square wave transform,
equipped with meaningful transform domain
(\mbox{cf.} the SDCT by Haweel~\cite{haweel2001square}).
It is important to recognize
that this relation
guarantees the perfect reconstruction property.

Additionally,
whenever a scaled version of the spectral components
is admissible,
the quantity $\beta$ (Equation~(\ref{eq.method.1.inv}))
can be set to one.
Possible scenarios are
Fourier descriptors evaluation~\cite{wonga2007shape},
transform domain threshold-based detectors~\cite{bellan2009sinusoids},
and
signal classification based on transform coefficient features~\cite{miao2002classification}.

\subsection{$8$- and $16$-point approximate DHT}

Besides being an elementary
short blocklength transform employed as
a means to evaluate larger size transforms~\cite{bracewell1996hartley},
the 8-point DHT has found applications as an operator for edge detection~\cite{park1998eight}.
The 1st order 8-point approximate DHT
also implies
another simple square wave transform,
according to:
\begin{align}
(\mathbf{H}_8^{(1)})^{-1}
=
\frac{1}{4}
(\mathbf{H}_8^{(0)}).
\end{align}
This fortunate relation
allows the definition of a real-valued transformation,
whose spectrum approximates the $8$-point DHT.

Now let us examine a fast algorithm for the
$0$th order 16-point approximate DHT.
The obtained transformation matrix $\mathbf{H}_{16}^{(0)}$ is given by
\begin{align}
\mathbf{H}_{16}^{(0)}
=
\left[
\begin{smallmatrix}
 1 &  1 &  1 &  1 &  1 &  1 &  1 &  1 &  1 &  1 &  1 &  1 &  1 &  1 &  1 & 1 \\
 1 &  1 &  1 &  1 &  1 &  1 &    &  - &  - &  - &  - &  - &  - &  - &    & 1 \\
 1 &  1 &  1 &    &  - &  - &  - &    &  1 &  1 &  1 &    &  - &  - &  - &   \\
 1 &  1 &    &  - &  - &  1 &  1 &  1 &  - &  - &    &  1 &  1 &  - &  - & - \\
 1 &  1 &  - &  - &  1 &  1 &  - &  - &  1 &  1 &  - &  - &  1 &  1 &  - & - \\
 1 &  1 &  - &  1 &  1 &  - &    &  1 &  - &  - &  1 &  - &  - &  1 &    & - \\
 1 &    &  - &  1 &  - &    &  1 &  - &  1 &    &  - &  1 &  - &    &  1 & - \\
 1 &  - &    &  1 &  - &  1 &  - &  1 &  - &  1 &    &  - &  1 &  - &  1 & - \\
 1 &  - &  1 &  - &  1 &  - &  1 &  - &  1 &  - &  1 &  - &  1 &  - &  1 & - \\
 1 &  - &  1 &  - &  1 &  - &    &  1 &  - &  1 &  - &  1 &  - &  1 &    & - \\
 1 &  - &  1 &    &  - &  1 &  - &    &  1 &  - &  1 &    &  - &  1 &  - &   \\
 1 &  - &    &  1 &  - &  - &  1 &  - &  - &  1 &    &  - &  1 &  1 &  - & 1 \\
 1 &  - &  - &  1 &  1 &  - &  - &  1 &  1 &  - &  - &  1 &  1 &  - &  - & 1 \\
 1 &  - &  - &  - &  1 &  1 &    &  - &  - &  1 &  1 &  1 &  - &  - &    & 1 \\
 1 &    &  - &  - &  - &    &  1 &  1 &  1 &    &  - &  - &  - &    &  1 & 1 \\
 1 &  1 &    &  - &  - &  - &  - &  - &  - &  - &    &  1 &  1 &  1 &  1 & 1
\end{smallmatrix}
\right].
\end{align}

\begin{figure}
\centering
\input{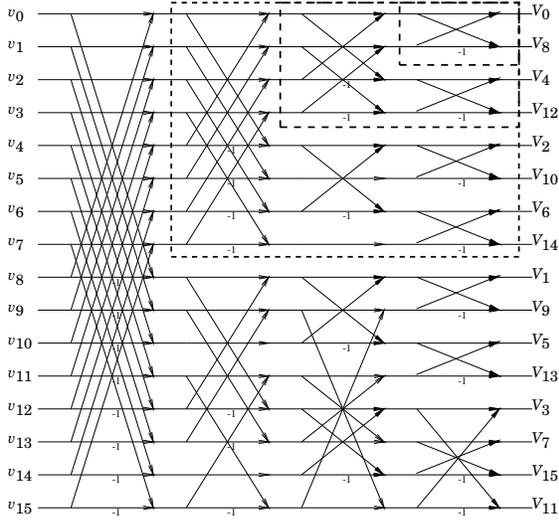}
\caption{Flow graph for the fast algorithm of $0$th order 16-point approximate DHT.
Dashed boxes denote shorter transforms embedded in 16-point algorithm.}
\label{fig:diagram}
\end{figure}

Using classical methods described in~\cite{blahut1985fast},
the implementation diagram of a fast algorithm for $\mathbf{H}_{16}^{(0)}$
was found
and is displayed in Figure~\ref{fig:diagram}.
The resulting algorithm turned out to possess embedding properties.
In the 16-point algorithm,
the presence
of fast algorithms
for the
$0$th order 2-, 4-, and 8-point approximate DHT
is noticeable.
In Figure~\ref{fig:diagram},
such smaller transformations are separated in dashed boxes.

\subsection{$8$-point approximate DCT}

The $8$-point DCT
has attracted considerable research effort.
This particular blocklength is
widely adopted in several image and video coding standards,
such as
JPEG, \mbox{MPEG-1}, \mbox{MPEG-2}, \mbox{H.261}, and \mbox{H.263}~\cite{roma2007hybrid}.
The $8$-point DCT is also subject to an extensive analysis in~\cite{britanak2007discrete}.
Using Equations~(\ref{eq.method.1.fwd}) and~(\ref{eq.method.2.fwd}),
we propose a new approximation for the $8$-point DCT.

In order to evaluate the suggested approximations,
we consider an input vector modelled
after
a first-order stationary Markov process with
zero mean and unity variance.
Additionally, it is assumed that adjacent vector components possess
a correlation coefficient of $0.95$~\cite{britanak2007discrete}.
Considering these assumptions,
commonly employed evaluation criteria, such as
(i) mean square error;
(ii) transform coding gain ($C_g$);
and
(iii) transform efficiency ($\eta$),
can be computed deterministically~\cite{liang2001fast}.

Tables~\ref{tab.rdct.scaling} and~\ref{tab.rdct.adjust}
list evaluation data for the proposed approximate
$8$-point DCT,
when
Equation~(\ref{eq.method.1.fwd}) and~(\ref{eq.method.2.fwd})
are considered, respectively.
High values for the coding gain in Table~\ref{tab.rdct.scaling}
are due to the
nonorthogonality of
the considered transformations.
This phenomenon was already reported in~\cite{raikar2003noise}.
In~\cite[p.~18]{goyal2001foundations},
Goyal gives a comprehensive account on how nonorthogonal transforms could
outperform the Kahunen-Lo\`{e}ve transform, for instance.

\begin{table}
\caption{Approximate DCT}
\label{tab.rdct.scaling}
\centering
\begin{tabular}{ccccc}
\noalign{\smallskip}
\hline
\noalign{\smallskip}
$m$ & $\beta$ & MSE & $C_g$ &  $\eta$ \\
\noalign{\smallskip}
\hline
\noalign{\smallskip}
0 & $0.3579$ & $1.1526\mathrm{e}{-2}$ & $9.9761$ & $90.2064$ \\
1 & $0.6502$ & $2.7587\mathrm{e}{-2}$ & $5.4135$ & $85.3438$ \\
2 & $0.4740$ & $1.3264\mathrm{e}{-3}$ & $9.1076$ & $92.0474$ \\
3 & $0.4741$ & $6.9859\mathrm{e}{-4}$ & $9.3152$ & $93.4396$ \\
4 & $0.5122$ & $1.5405\mathrm{e}{-4}$ & $8.5446$ & $93.4021$ \\
5 & $0.4927$ & $4.2514\mathrm{e}{-5}$ & $8.9304$ & $94.1522$ \\
6 & $0.5025$ & $1.0982\mathrm{e}{-5}$ & $8.7841$ & $93.8968$ \\
7 & $0.4976$ & $4.4935\mathrm{e}{-6}$ & $8.8745$ & $93.9519$ \\
$\infty$
  & $1/2$    & --                     & $8.8259$ & $93.9912$ \\
\noalign{\smallskip}
\hline
\end{tabular}
\end{table}

\begin{table}
\caption{Approximate DCT using $\mathbf{S}^{-1}$}
\label{tab.rdct.adjust}
\centering
\begin{tabular}{ccccc}
\noalign{\smallskip}
\hline
\noalign{\smallskip}
$m$ &  MSE & $C_g$ &  $\eta$ \\
\noalign{\smallskip}
\hline
\noalign{\smallskip}
0 & $9.8002\mathrm{e}{-3}$ & $8.1827$ & $87.4297$ \\
1 & $4.6128\mathrm{e}{-4}$ & $8.8007$ & $92.8519$ \\
2 & $5.9758\mathrm{e}{-4}$ & $8.7760$ & $92.1149$ \\
3 & $3.2740\mathrm{e}{-4}$ & $8.7880$ & $92.9931$ \\
4 & $4.0163\mathrm{e}{-5}$ & $8.8201$ & $93.5807$ \\
5 & $2.0875\mathrm{e}{-5}$ & $8.8262$ & $94.0782$ \\
6 & $6.6373\mathrm{e}{-6}$ & $8.8249$ & $93.8981$ \\
7 & $1.1358\mathrm{e}{-6}$ & $8.8254$ & $93.9244$ \\
$\infty$
  & --                     & $8.8259$ & $93.9912$ \\
\noalign{\smallskip}
\hline
\end{tabular}
\end{table}

Existing approximation methods for the DCT
include
(i) the $C$-matrix transform (CMT)~\cite{jones1978related};
(ii) the integer cosine transform (ICT)~\cite{cham1987integer};
(iii) the generalized Chen transform (GCT)~\cite{allen1991chen};
and
(iv) the binDCT algorithm~\cite{tran2000bindct}.
The $C$-matrix transform consists of an approximation
for the $8$-point DCT transform
using the Walsh-Hadamard transform as a pre-processing stage,
followed by a conversion matrix of integer entries~\cite{jones1978related}.
The integer cosine transform adopts a somewhat different approach.
It directly
approximates the DCT matrix by integer elements
without any pre-processing.

The GCT takes advantage of a parametrization of
the DCT matrix replacing exact parameter values by rational approximations,
such as $3/16,3/8,11/16,91/128$~\cite[p.~211]{britanak2007discrete}.
Being the parameters multiplicatively combined,
the GCT scheme can provide a final approximate DCT matrix
with elements of large integer representation (e.g., $1729/2048$ or $1183/2048$).
In its turn,
the binDCT employs an approach
based on lifting schemes~\cite{liang2001fast}.
Although the individual multiplicative elements of the
binDCT lifting structure are relatively small (e.g., $13/16$ or $15/16$),
they are multiplied in cascade.
The resulting basis vectors that approximate
the
DCT matrix
possess
elements such as $7823/8192$ or $3217/4096$~\cite[p.~229]{britanak2007discrete}.
So the elements of the
final
effective transformation matrix have
a significantly larger dynamic range
when compared to that of
the individual constants employed by
GCT parametrization or
by
the
binDCT lifting scheme.

Therefore,
the final approximate quantities could be taken into consideration
when deriving a comparison between approximation methods.
Since the accuracy of approximation is closely related
to the dynamic range of the utilized integer numbers,
a fair comparison of performance
could limit the size of the considered bit representation
in the final approximate matrix.
Table~\ref{tab.comparison.1}
brings a quantitative comparison of the suggested methodology
with the referred alternative methods
described in literature.
For each method,
the dynamic range of the final approximation matrix is shown in parenthesis;
design parameters are also indicated for the proposed methodology.

\begin{table*}
\caption{Comparison with some existing methods}
\label{tab.comparison.1}
\centering
\begin{tabular}{lccc}
\noalign{\smallskip}
\hline
\noalign{\smallskip}
Method & MSE & $C_g$ &  $\eta$ \\
\noalign{\smallskip}
\hline
\noalign{\smallskip}
$\mathrm{ICT}_8$-II ($3$-bit) %
             & $2.7217\mathrm{e}{-3}$ & $8.6513$ & $91.1212$ \\
Proposed ($3$-bit, $m=2$, $\beta = 1/2$)     & $4.3366\mathrm{e}{-3}$ & $8.8755$ & $92.0474$ \\
Proposed ($3$-bit, $m=2$, $\beta = 0.4831$)  & $1.6963\mathrm{e}{-3}$ & $9.0248$ & $92.0474$ \\
\noalign{\smallskip}
\hline
\noalign{\smallskip}
$\mathrm{CMT}_8$ ($4$-bit)        %
             & $3.3003\mathrm{e}{-3}$ & $8.8259$ & $93.9912$ \\
$\mathrm{ICT}_8$-II ($4$-bit) %
             & $2.0607\mathrm{e}{-4}$ & $8.8141$ & $94.0945$ \\
Proposed ($4$-bit, $m=3$, $\beta = 1/2$)     & $3.6916\mathrm{e}{-3}$ & $9.0840$ & $93.4396$ \\
Proposed ($4$-bit, $m=3$, $\beta = 0.4925$)  & $2.2113\mathrm{e}{-3}$ & $9.1496$ & $93.4396$ \\
\noalign{\smallskip}
\hline
\noalign{\smallskip}
$\mathrm{ICT}_8$-II ($5$-bit) %
             & $1.3029\mathrm{e}{-4}$ & $8.8144$ & $93.9799$ \\
binDCT-IIC ($13$-bit)
             & $2.7190\mathrm{e}{-4}$ & $8.8160$ & $93.0667$ \\
Proposed ($5$-bit, $m=5$, $\beta = 1/2$)     & $2.6404\mathrm{e}{-4}$ & $8.8664$ & $94.1521$ \\
Proposed ($5$-bit, $m=5$, $\beta = 0.4966$)  & $1.0624\mathrm{e}{-4}$ & $8.8961$ & $94.1522$ \\
\noalign{\smallskip}
\hline
\noalign{\smallskip}
$\mathrm{CMT}_8$ ($6$-bit)        %
             & $2.2680\mathrm{e}{-4}$ & $8.8070$ & $93.1832$ \\
Proposed ($6$-bit, $m=6$, $\beta = 1/2$)     & $3.5499\mathrm{e}{-5}$ & $8.8053$ & $93.8968$ \\
Proposed ($6$-bit, $m=6$, $\beta = 0.5009$)  & $2.0968\mathrm{e}{-5}$ & $8.7975$ & $93.8968$ \\
\noalign{\smallskip}
\hline
\noalign{\smallskip}
$\mathrm{GCT}_8$-II ($11$-bit)        %
             & $4.2336\mathrm{e}{-5}$ & $8.8226$ & $93.9912$ \\
Proposed ($7$-bit, $m=7$, $\beta = 1/2$)     & $2.7188\mathrm{e}{-5}$ & $8.8539$ & $93.9519$ \\
Proposed ($7$-bit, $m=7$, $\beta = 0.4995$)  & $1.8626\mathrm{e}{-5}$ & $8.8582$ & $93.9519$ \\
\noalign{\smallskip}
\hline
\end{tabular}
\end{table*}

The low complexity of the proposed $0$th order approximate DCT
can be of additional practical interest;
$\mathbf{C}_8^{(0)}$
requires only 24 additions.
Moreover,
its orthogonalizing adjustment matrix is a simple diagonal matrix
given by:
\begin{align}
\mathbf{S}^{-1}
=
\mathrm{diag}%
\left(
\frac{1}{\sqrt{8}},
\frac{1}{\sqrt{6}},
\frac{1}{2},
\frac{1}{\sqrt{6}},
\frac{1}{\sqrt{8}},
\frac{1}{\sqrt{6}},
\frac{1}{2},
\frac{1}{\sqrt{6}}
\right)
.
\end{align}
Considering the adjustment offered by $\mathbf{S}^{-1}$,
the final approximation offers a MSE of $9 \times 10^{-3}$.
For comparison,
several versions of the $\mathrm{ICT}_8$-II
could only exhibit a similar MSE performance,
ranging from
$6 \times 10^{-3}$ to $8 \times 10^{-3}$~\cite[p.~177]{britanak2007discrete},
at the expense of using
$2$- or $3$-bit arithmetic.

If the proposed nonorthogonal formulation that requires
$\mathbf{D}^{-1}$ is chosen
(Equations~(\ref{eq.method.1.fwd}) and (\ref{eq.method.1.inv})),
we obtain that
\begin{align}
(\mathbf{C}_8^{(0)})^{-1}
=
(\mathbf{C}_8^{(0)})^H
\mathbf{D}^{-1},
\end{align}
where
$\mathbf{D}^{-1} =
\mathrm{diag}(1/8, 1/6, 1/4, 1/6, 1/8, 1/6, 1/4, 1/6)$.
The diagonal elements are small and impose low computational requirements.
This can be interpreted as the basis relation for the definition of another new square wave transform.

Additionally,
in any case,
when considering the DCT as a pre-processing step for
a subsequent coefficient quantization procedure for image compression,
the elements of $\mathbf{D}^{-1}$ can be included into the quantization step.
This procedure is suggested and adopted in several works~\cite{bouguezel2008low,lengwehasatit2004scalable}.
As a consequence,
the computational complexity
of
the approximation is totally confined to that of $\mathbf{C}_8^{(0)}$.

\begin{figure*}
\centering
\subfigure[]{\epsfig{file=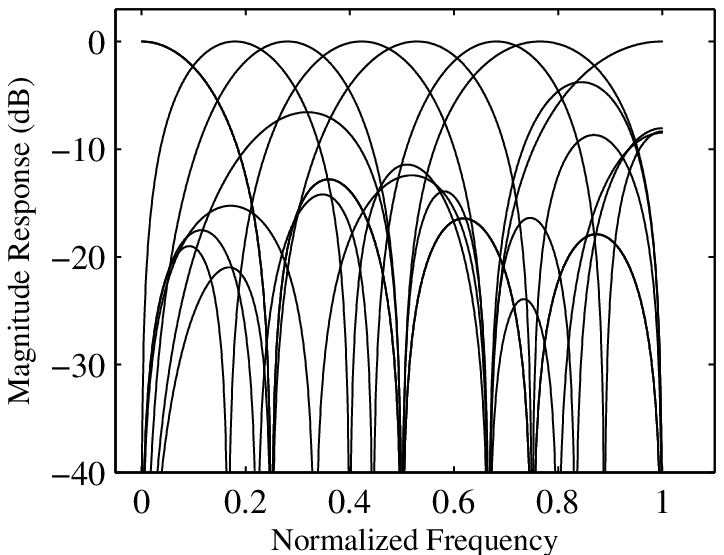,width=7cm}}
\subfigure[]{\epsfig{file=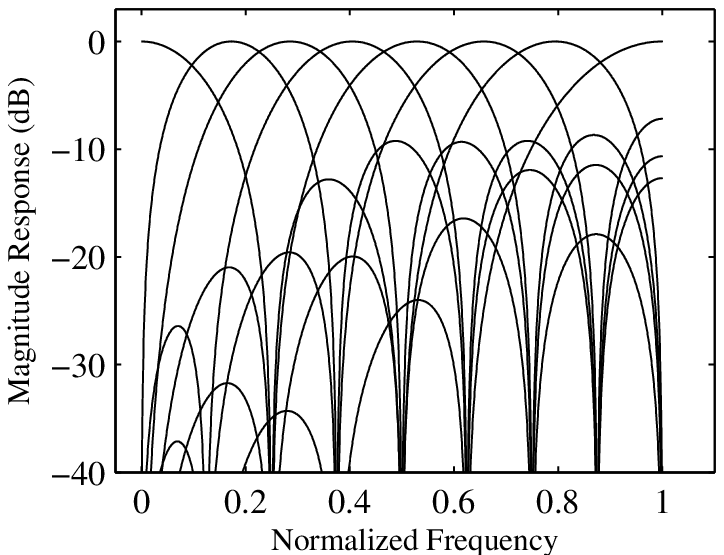,width=7cm}}
\caption{Fourier transform (magnitude) of the basis vectors
of
(a) $\mathbf{C}_8^{(0)}$
and
(b) the exact DCT.}
\label{fig.freq.response}
\end{figure*}

Additionally,
Figure~\ref{fig.freq.response}(a)
illustrates the absence of DC leakage of
$\mathbf{C}_8^{(0)}$.
All frequency response curves vanish
at null frequency,
except, of course,
the first curve,
which corresponds to the
moving average filter associated
to the first row of the transformation matrix.
Figure~\ref{fig.freq.response}(b)
shows the frequency response of the exact DCT for comparison~\cite{strang1999dct}.

\section{Conclusions}
\label{sec.conclusions}

Using a simple and straightforward approach,
we demonstrate that several approximate transforms can be
conveniently obtained.
It is important to emphasize that the proposed methods are general approaches,
encompassing distinct transforms and various blocklengths.
This fact contrasts with the highly specialized procedures
archived in literature.
Nevertheless,
we could still derive meaningful performance comparisons.

The proposed integer approximations
are well suited for
architectures that take advantage of the
dyadic rationals and
canonical signed digit representation.
Overall,
low computational complexities
are obtained.
Further dedicated optimization methods could enhance
the proposed methods for selected blocklengths and kernels.
Possible venues include the elaboration of specially designed
algorithms
for the computation of particular adjustment matrices.
Additionally,
new $8$-point square wave transforms
equipped with perfect reconstruction
and meaningful spectra
were suggested
for the DFT, DHT, and DCT.

\section*{Acknowledgments}

This work was partially supported
by
the Department of Foreign Affairs and International Trade of Canada
and
the
\emph{Conselho Nacional de Desenvolvimento Cient\'ifico e Tecnol\'ogico} (CNPq), Brazil.

{\small
\singlespacing
\bibliographystyle{ieeetr}
\bibliography{ref}
}

\end{document}